\documentclass[preprint,showpacs,preprintnumbers,amsmath,amssymb,floatfix,endfloats*]{revtex4}
\usepackage{graphicx}% Include figure files
\usepackage{dcolumn}% Align table columns on decimal point
\usepackage{bm}% bold math
\usepackage{amsfonts}

\DeclareMathAlphabet{\mathsfsl}{OT1}{cmr}{bx}{it}
\begin{document}
%----------------------------------------------------------------------%
% Title
%----------------------------------------------------------------------%
\title{Shear rate threshold for the boundary slip in dense polymer films}
\author{Nikolai~V.~Priezjev}

\affiliation{Department of Mechanical Engineering, Michigan State
University, East Lansing, Michigan 48824}
\date{\today}
%
%----------------------------------------------------------------------%
%\twocolumn[
%----------------------------------------------------------------------%
% Abstract
%----------------------------------------------------------------------%
\begin{abstract}

The shear rate dependence of the slip length in thin polymer films
confined between atomically flat surfaces is investigated by
molecular dynamics simulations. The polymer melt is described by the
bead-spring model of linear flexible chains. We found that at low
shear rates the velocity profiles acquire a pronounced curvature
near the wall and the absolute value of the negative slip length is
approximately equal to thickness of the viscous interfacial layer.
At higher shear rates, the velocity profiles become linear and the
slip length increases rapidly as a function of shear rate. The
gradual transition from no-slip to steady-state slip flow is
associated with faster relaxation of the polymer chains near the
wall evaluated from decay of the time autocorrelation function of
the first normal mode. We also show that at high melt densities the
friction coefficient at the interface between the polymer melt and
the solid wall follows power law decay as a function of the slip
velocity. At large slip velocities the friction coefficient is
determined by the product of the surface induced peak in the
structure factor, temperature and the contact density of the first
fluid layer near the solid wall.

\end{abstract}

\pacs{68.08.-p, 83.80.Sg, 83.50.Rp, 47.61.-k, 83.10.Rs}

%   83.50.Rp  Wall slip and apparent slip
%   83.80.Sg  Polymer melts
%   47.61.-k  Micro- and nano- scale flow phenomena
%   68.08.-p  Liquid–solid interfaces
%   83.10.Rs  Computer simulation of molecular and particle dynamics

\maketitle

\section{Introduction}

\begin{figure}[t]
\vspace*{-3mm}
\includegraphics[width=9.0cm,angle=0]{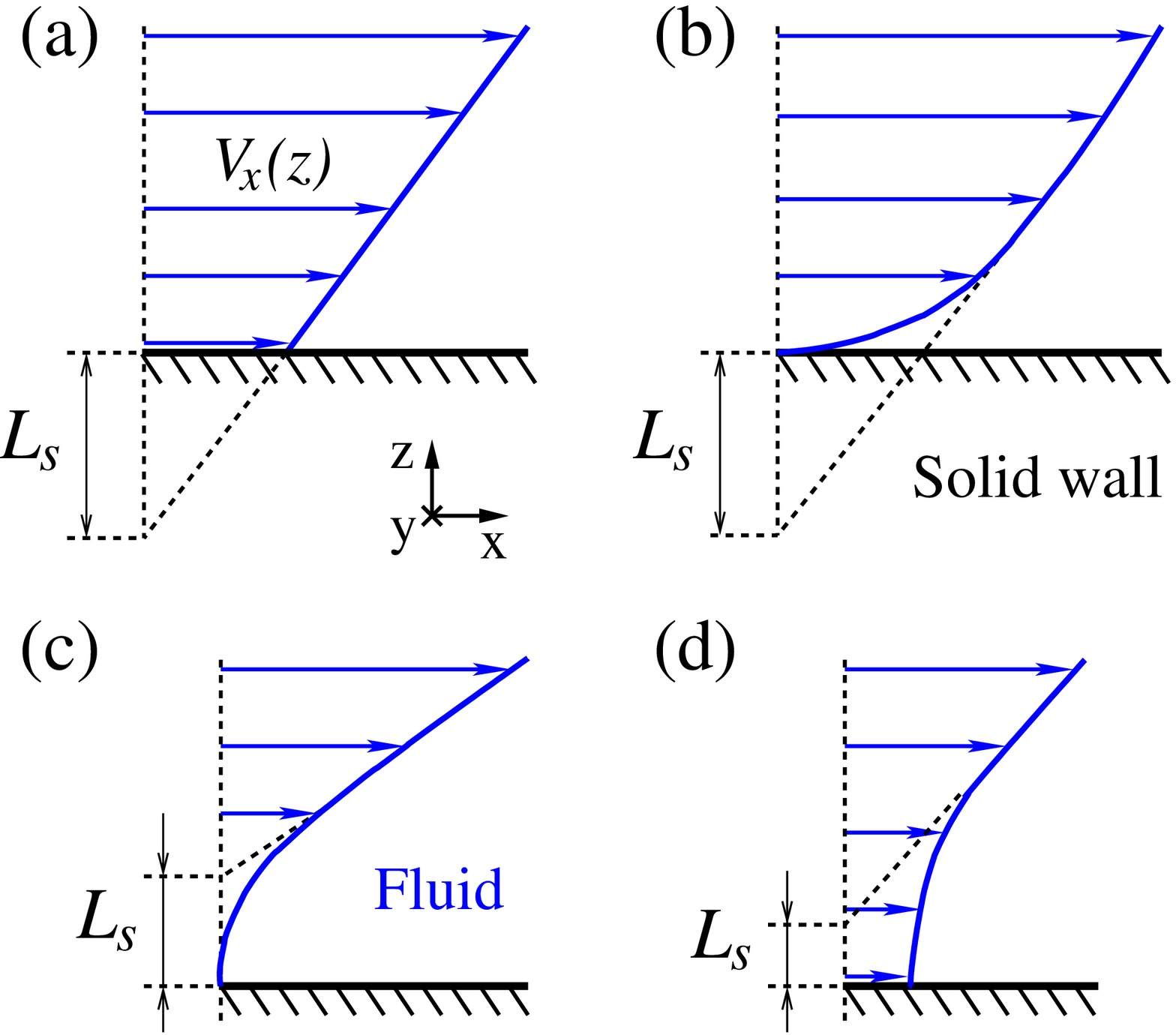}
\caption{(Color online) A schematic diagram of a steady-state shear
flow over a flat solid wall. (a) The slip velocity and the slope of
the linear velocity profile are related via
$V_s\,{=}\,\,\dot{\gamma}L_s$, where $L_s$ is the slip length. (b)
Apparent slip is associated with the lower viscosity boundary layer.
(c) The velocity profile with a downward curvature due to higher
viscosity boundary layer is described by the negative slip length.
(d) A combination of the `true' slip at the liquid/solid interface
and the curvature of the velocity profile.} \label{schematic}
\end{figure}

The rheology of complex fluids in thin films is important for
theoretical and experimental studies of such common phenomena as
friction, lubrication and wear~\cite{Mate08}. Experimental
measurements of the flow profiles and shear stresses on submicron
scales might be subject to errors due to the possibility of liquid
slip at the solid wall. An accurate prediction of flow, therefore,
requires specification of a proper boundary condition. In the Navier
model the interfacial shear rate and slip velocity are related via
the proportionality coefficient, the so-called \textit{slip length}.
When the adjacent fluid layer slides past a solid wall with a finite
velocity, the slip length is computed by linear extrapolation of the
velocity profile near the interface to zero velocity [see
Fig.\,\ref{schematic}\,(a)].  In the case when there is no relative
velocity between fluid and solid at the interface, the formation of
a lower viscosity boundary layer might result in the apparent slip
length which is defined by the slope of the bulk velocity
profile~\cite{VinogradLang95}~[see Fig.\,\ref {schematic}\,(b)].
Experimental studies on pressure-driven flows in microchannels or
thin film drainage using either the surface force apparatus (SFA) or
atomic force microscope (AFM) have demonstrated that the slip length
depends on the nanoscale surface
roughness~\cite{Granick02,Archer03,Leger06,Vinograd06}, surface
wettability~\cite{Churaev84,Charlaix01,CottinEPJE02}, rate of
shear~\cite{Granick01,Granick02,CraigPRL01,GranLang02,Breuer03,Ulmanella08},
and fluid structure~\cite{MackayVino,SchmatkoPRL05}. Several slip
regimes exist due to progressive disentanglement of the anchoring
chains in the shear flow of polymer
melts~\cite{deGennes92,LegerPRL93}. The difficulty in experimental
determination of the velocity profiles and structure of complex
fluids near interfaces leaves open important questions regarding the
shear rate dependency of the slip length and the existence of a
shear rate threshold for the boundary slip.

In the last two decades, the boundary conditions at the interface
between monatomic liquids and atomically flat walls were extensively
studied by molecular dynamics (MD)
simulations~\cite{Fischer89,KB89,Thompson90,Barrat99,Barrat99fd,Cieplak01,Quirke01,Attard04,Priezjev07,PriezjevJCP}.
The main factors affecting the slip are the energy of wall-fluid
interaction and commensurability of the liquid and solid structures
at the interface. At high wall-fluid energies, the first layer of
fluid monomers becomes epitaxially locked to the solid substrate and
the effective no-slip boundary plane is displaced into the fluid
region. The absolute value of the negative slip length is
approximately equal to the number of stacked monolayers between the
effective and real boundary planes~\cite{Fischer89,Thompson90}. At
low surface energy, the first layer can slide with a finite velocity
relative to the solid substrate under the shear stress from the bulk
fluid [this situation is sometimes referred to as a molecular or
`true' slip, e.g., see Fig.\,\ref{schematic}\,(a)]. The slip length
is inversely proportional to the peak of the in-plane structure
factor computed in the first fluid layer at the main reciprocal
lattice vector~\cite{Thompson90,Barrat99fd,Priezjev07}. The exact
relation between the slip length and microscopic parameters of the
liquid/solid interface, however, has not yet been established.

The variation of the slip length with increasing shear rate in the
flow of simple fluids past atomically smooth walls was first
reported in the MD study by Thompson and Troian~\cite{Nature97}. The
nonlinear rate dependence of the slip length was well fitted by a
power law function for different wall densities and weak wall-fluid
interaction energies~\cite{Nature97}. In a later
study~\cite{Priezjev07}, it was shown that the slip length is a
linear function of shear rate at high wall-fluid interaction
energies and, when the surface energy is reduced, the
rate-dependence of the slip length can be well fitted by the power
law function proposed in Ref.\,\cite{Nature97}. It was also found
that in a wide range of shear rates and wall-fluid interaction
energies, the slip length is a function of a single variable which
combines temperature of the first fluid layer, the contact density,
and the peak value of the in-plane fluid structure factor evaluated
at the main reciprocal lattice vector~\cite{Priezjev07}. The results
of previous MD studies of monatomic fluids lead to a conclusion that
the boundary conditions for flows past smooth surfaces are either
no-slip and rate-independent~\cite{Thompson90} or described by the
finite positive slip length which increases with shear
rate~\cite{Nature97,Priezjev07,PriezjevJCP}. Except for flows of low
density fluids~\cite{Fang05}, there was no reported observation of a
transition from no-slip to steady-state slip flow with increasing
shear rate for simple fluids.

The slip length in a flow of a polymer melt past a flat passive
(non-adsorbing) surface is a ratio of the fluid viscosity to the
friction coefficient which is determined by the interaction between
fluid monomers and wall atoms~\cite{deGennes85}. In a recent MD
study of unentangled melts confined between smooth
surfaces~\cite{Priezjev04}, it was found that the friction
coefficient at the liquid/solid interface is nearly independent of
the chain length beyond ten bead-spring units and, therefore, the
molecular weight dependence of the slip length at low shear rates is
mostly dominated by the melt viscosity. Depending on the strength of
the wall-fluid interaction energy, the slip occurs either at the
confining
surfaces~\cite{Manias93,Thompson95,Manias96,dePablo96,StevensJCP97,Koike98,Tanner99,Priezjev04,Priezjev08,Servantie08,Pastorino08,Niavarani08}
or between the adsorbed layer and free polymer
chains~\cite{Manias93,Manias96}. A transition from stick to slip
flow with increasing shear rate was reported in MD simulations of
thin films of hexadecane~\cite{Tanner99} and
oligomers~\cite{Manias96} confined between strongly adsorbing
surfaces. Despite extensive efforts in molecular simulations of thin
polymer films, it still remains unclear what system parameters
(fluid density, pressure, chain length, surface energy) determine
the onset of boundary slip at the interface between an unentangled
melt and a flat surface.

In a previous MD study~\cite{Priezjev08}, the effect of shear rate
on slip boundary conditions in thin polymer films confined between
atomically smooth surfaces was investigated as a function of melt
density. It was found that the slip length, extracted from the
linear velocity profiles, passes through a local minimum at low
shear rates and then increases rapidly at higher shear rates. This
behavior was rationalized in terms of the friction coefficient
(defined as a ratio of the shear-thinning viscosity to the slip
length), which undergoes a gradual transition from a nearly constant
value to the power law decay as a function of the slip velocity. The
functional form of the relation between the friction coefficient and
the slip velocity in the crossover region can also be considered as
a boundary condition for monatomic
fluids~\cite{Priezjev08,Nature97}. In addition, it was shown that in
a wide range of shear rates and melt densities, the friction
coefficient is determined by the product of the value of surface
induced peak in the structure factor, temperature and the contact
density of the first fluid layer near the solid
wall~\cite{Priezjev08}. Whether these conclusions remain valid at
higher melt densities (when a viscous interfacial layer is formed
near the confined surfaces) is one of the motivations of the present
study.

In this paper, we investigate the rate dependence of the slip length
at the interface between an unentangled polymer melt and atomically
flat walls using molecular dynamics simulations. We will show that
in dense polymer films the velocity profiles are curved at low shear
rates [shown schematically in Fig.\,\ref{schematic}\,(c) and
Fig.\,\ref{schematic}\,(d)] due to highly viscous interfacial layer
and the corresponding slip length is negative. When the viscosity of
the interfacial layer is reduced at higher shear rates, the velocity
profiles become linear and the slip length increases rapidly as a
function of shear rate. The relaxation dynamics of polymer chains
near the walls and in the bulk region is studied by analyzing the
time autocorrelation function of the first normal mode at
equilibrium and in shear flow. We will also show that the friction
coefficient follows power law decay as a function of the slip
velocity and its universal dependence on microscopic parameters of
the liquid/solid interface holds only at large slip velocities.

The rest of this paper is organized as follows. The details of
molecular dynamics simulations and the equilibration procedure are
described in the next section. The fluid velocity and density
profiles, shear rate dependence of the slip length, and the analysis
of the fluid structure and relaxation dynamics of polymer chains are
presented in Section~\ref{sec:Results}. The conclusions are given in
the last section.

\section{Molecular dynamics simulation model}
\label{sec:Model}

\begin{figure}[t]
\vspace*{-3mm}
\includegraphics[width=10.0cm,angle=0]{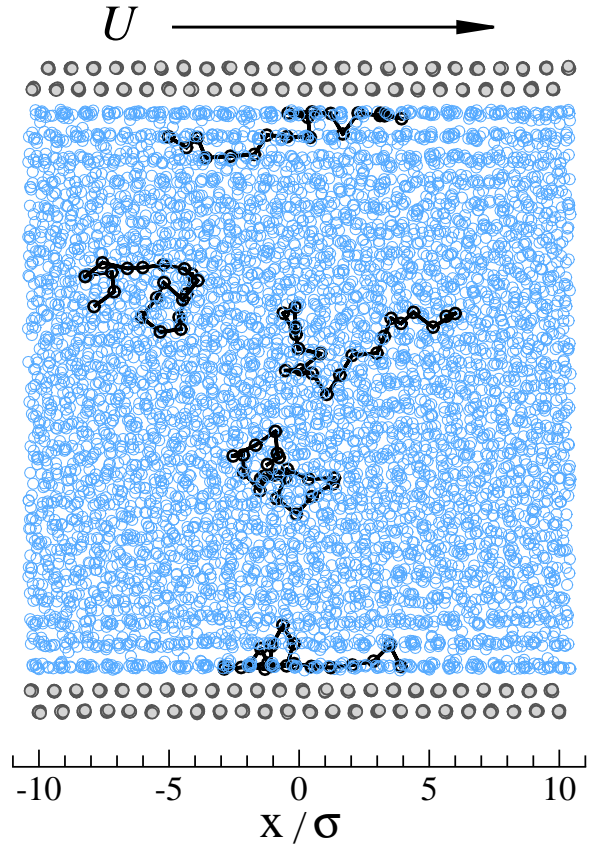}
\caption{(Color online) A snapshot of fluid monomers (open blue
circles) and wall atoms (filled gray circles) positions projected on
the $xz$ plane. The bottom wall is at rest and the top wall is
moving with a constant velocity $U$ in the $\hat{x}$ direction. Each
fluid monomer belongs to a polymer chain. Five polymer chains are
marked by solid lines. The fluid monomer density is
$\rho\,{=}\,1.08\,\sigma^{-3}$ and the top wall speed is
$U\,{=}\,\,0.1\,\sigma/\tau$.} \label{snapshot}
\end{figure}

The simulation setup is similar to that described in the previous
study~\cite{Priezjev08} of a polymeric fluid undergoing planar shear
flow between two atomically smooth walls. Figure\,\ref{snapshot}
shows a snapshot of an unentangled polymer melt confined between
solid walls. The total number of fluid monomers is kept the same as
in the previous study ($N_{f}\!=6000$) but the simulations are
performed at higher fluid densities. The fluid monomers interact via
the truncated Lennard-Jones (LJ) potential
\begin{equation}
V_{LJ}(r)\!=4\,\varepsilon\,\Big[\Big(\frac{\sigma}{r}\Big)^{12}\!-\Big(\frac{\sigma}{r}\Big)^{6}\,\Big]
~~\text{for}~r\leqslant r_c\!=\!2.5\,\sigma,
\end{equation}
where $\varepsilon$ and $\sigma$ are the energy and length scales of
the fluid phase. The interaction between the wall atoms and fluid
monomers is also modeled by the LJ potential with $\varepsilon_{\rm
wf}\,{=}\,0.9\,\varepsilon$ and $\sigma_{\rm wf}\,{=}\,\sigma$. The
wall atoms are tethered about the sites of an fcc lattice and do not
interact with each other.

The coarse-grained bead-spring model was used to represent an
unentangled polymer melt with linear chains of $N\,{=}\,20$
monomers. In addition to the LJ potential any two neighboring
monomers in the chain interact through the finitely extensible
nonlinear elastic (FENE) potential~\cite{Bird87}
\begin{equation}
V_{FENE}(r)=-\frac{k_s}{2}\,r_{\!o}^2\ln[1-r^2/r_{\!o}^2],
\end{equation}
with the standard parameters $k_s\,{=}\,30\,\varepsilon\sigma^{-2}$
and $r_{\!o}\,{=}\,1.5\,\sigma$~\cite{Kremer90}. A combination of
the LJ and FENE potentials between neighboring monomers yields an
effective spring potential, which is strong enough to prevent
polymer chains from unphysical crossing each other or bond breaking
even at the highest shear rates considered in the present study.

The dynamics of fluid molecules and wall atoms was weakly coupled to
a heat bath through a Langevin thermostat~\cite{Grest86}. The
thermostat was applied only to the $\hat{y}$ component
(perpendicular to the plane of shear) of the equations of motion for
fluid monomers to avoid a bias in the shear flow
direction~\cite{Thompson90}. The equations of motion for fluid
monomers in the $\hat{x}$, $\hat{y}$ and $\hat{z}$ directions are
given by
\begin{eqnarray}
\label{Langevin_x}
m\ddot{x}_i & = & -\sum_{i \neq j} \frac{\partial V_{ij}}{\partial x_i}\,, \\
\label{Langevin_y}
m\ddot{y}_i + m\Gamma\dot{y}_i & = & -\sum_{i \neq j} \frac{\partial V_{ij}}{\partial y_i} + f_i\,, \\
\label{Langevin_z}
m\ddot{z}_i & = & -\sum_{i \neq j} \frac{\partial V_{ij}}{\partial z_i}\,, %
\end{eqnarray}
where the sum is taken over the neighboring fluid monomers and wall
atoms within the cutoff radius $r_c\!=\!2.5\,\sigma$,
$\Gamma\,{=}\,1.0\,\tau^{-1}$ is the friction coefficient, and $f_i$
is a random uncorrelated force with zero mean and variance $\langle
f_i(0)f_j(t)\rangle\,{=}\,\,2mk_BT\Gamma\delta(t)\delta_{ij}$
determined from the fluctuation-dissipation relation. The
temperature of the Langevin thermostat is set to
$T\,{=}\,1.1\,\varepsilon/k_B$, where $k_B$ is the Boltzmann
constant. The equations of motion were integrated using the
fifth-order gear-predictor algorithm~\cite{Allen87} with a time step
$\triangle t\,{=}\,0.002\,\tau$, where
$\tau\!=\!\sqrt{m\sigma^2/\varepsilon}$ is the characteristic time
of the LJ potential. The relatively small time step was chosen to
resolve accurately the dynamics of fluid molecules and wall atoms at
the interface.

The fluid is confined by the flat solid walls in the $\hat{z}$
direction (see Fig.\,\ref{snapshot}). Each wall consists of $576$ LJ
atoms arranged in two layers of an fcc crystal with (111) plane
parallel to the $xy$ plane. The nearest-neighbor distance between
fcc lattice sites in the $xy$ plane is $d\,{=}\,1.0\,\sigma$ and the
wall density is $\rho_w\,{=}\,1.40\,\sigma^{-3}$. The wall atoms
were allowed to oscillate about their equilibrium lattice positions
under the harmonic potential $V_{sp}\,{=}\,\frac{1}{2}\,\kappa\,r^2$
with the spring stiffness $\kappa\,{=}\,1200\,\varepsilon/\sigma^2$.
The mean-square displacement of the wall atoms satisfies the
Lindemann criterion for melting, i.e., $\langle\delta
u^2\rangle/d^{\,2}\lesssim 0.023$. In addition, the random force and
the damping term were applied to all three components of the wall
atom equations of motion, e.g., for the $\hat{x}$ component
\begin{eqnarray}
\label{Langevin_wall_x} m_w\,\ddot{x}_i + m_w\,\Gamma\dot{x}_i & = &
-\sum_{i \neq j} \frac{\partial V_{ij}}{\partial x_i} -
\frac{\partial V_{sp}}{\partial x_i} + f_i\,,
\end{eqnarray}
where the mass of the wall atoms is $m_w\,{=}\,10\,m$, the friction
coefficient is $\Gamma\,{=}\,1.0\,\tau^{-1}$ and the sum is taken
over the fluid monomers within the cutoff distance
$r_c\!\,\,{=}\,\,2.5\,\sigma$. Periodic boundary conditions were
applied in the $\hat{x}$ and $\hat{y}$ directions parallel to the
walls.

The polymer melt was initially equilibrated for about
$5\times10^4\,\tau$ at a constant normal pressure
$P=0.5\,\varepsilon\,\sigma^{-3}$ applied on the upper wall while
the lower wall was kept at rest. Then, the external pressure was
gradually increased to a desired value (reported in
Table\,\ref{tabela}). The distance between the walls was fixed after
the system was additionally equilibrated at the constant pressure
for about $2\times10^4\,\tau$. The MD simulations described below
were performed at a constant density ensemble. The fluid density,
the corresponding pressure in the absence of shear flow, and the
channel height are listed in Table\,\ref{tabela}.

The shear flow was generated by moving the upper wall with a
constant velocity $U$ in the $\hat{x}$ direction parallel to the
stationary lower wall. Before the production run started, the flow
was simulated for about $5\times10^4\tau$ for each value of the
upper wall speed. The fluid velocity and density profiles were
averaged within bins of thickness $\Delta z\,{=}\,0.01\,\sigma$ for
a time period up to $6\times10^5\tau$ at the lowest upper wall speed
$U\,{=}\,\,0.001\,\sigma/\tau$. The small bin size was chosen to
resolve accurately the velocity profiles and fluid structure near
the walls. At the highest shear rates examined in this study, the
Reynolds number did not exceed the value $Re\approx1.8$ estimated
from the maximum difference of the fluid velocities near the upper
and lower walls, the shear viscosity, the fluid density and the
channel height (see Table\,\ref{tabela}).

\begin{table}[b]
\caption{The fluid density $\rho$ and pressure $P$ at equilibrium
($U=0$). The density is defined as a ratio of the total number of
fluid monomers ($N_{f}\!=6000$) to the volume $20.86\,\sigma \times
12.04\,\sigma \times (h-\sigma)$, where $h$ is the distance between
the fcc lattice planes in contact with the fluid.}
 \vspace*{3mm}
 \begin{ruledtabular}
 \begin{tabular}{r r r r r r}
   $\rho\,(\text{units of} \,\,\sigma^{-3})$ &            $1.04$ & $1.06$ & $1.08$ & $1.09$ & $1.11$
   \\ [3pt] \hline %\\[-5pt]
   $P\,(\text{units of}\,\,\,\varepsilon\,\sigma^{-3})$ & $6.0$ &  $7.0$ &  $8.0$ &  $9.0$ &  $10.0$
   \\ [3pt]
   $h\,(\text{units of} \,\,\sigma)$ &                  $24.01$ & $23.50$ & $23.10$ & $22.83$ & $22.54$
   \\ [2pt]
 \end{tabular}
 \end{ruledtabular}
 \label{tabela}
\end{table}

\section{Results}
\label{sec:Results}

\subsection{Fluid density and velocity profiles}

Molecular simulations of polymer melts confined by flat walls have
demonstrated that the fluid structure consists of several discrete
layers of monomers, the polymer chains are flattened near the walls,
and the chain configurations remain bulklike in a region of several
molecular diameters away from the walls~\cite{Bitsanis90}.
Qualitatively, these features can be observed in the snapshot of the
polymer film shown in Fig.\,\ref{snapshot}. Note that the first two
layers of monomers near the walls are clearly distinguishable, the
polymer chains in contact with the wall atoms are compressed towards
the interface and their constitutive monomers are located within the
discrete layers.

\begin{figure}[t]
\includegraphics[width=12.cm,angle=0]{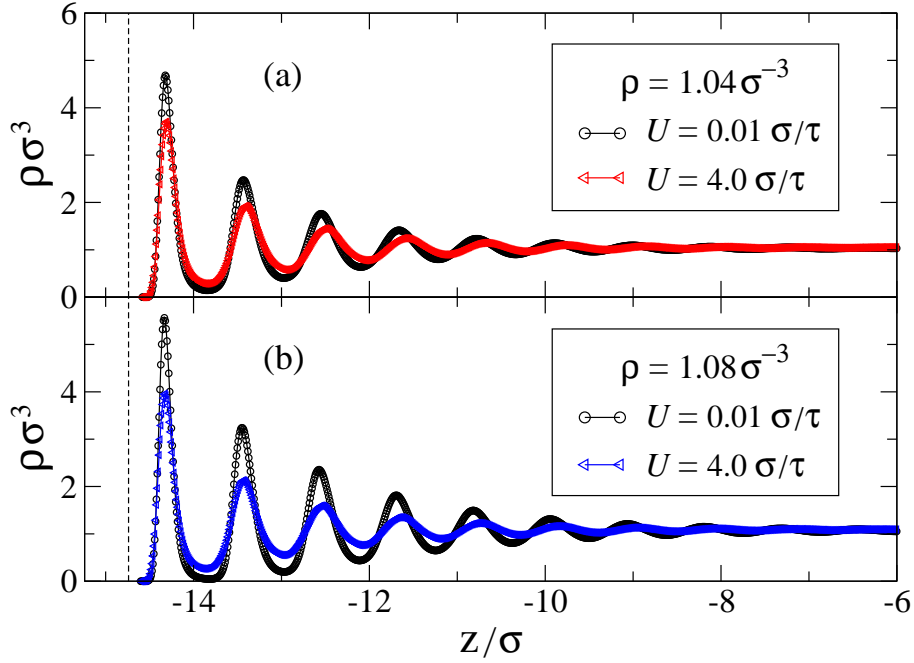}
\caption{(Color online) Averaged monomer density profiles near the
stationary lower wall with $\varepsilon_{\rm
wf}/\varepsilon\,{=}\,0.9$. The uniform fluid density away from the
walls is $\rho\,{=}\,1.04\,\sigma^{-3}$ (a) and
$\rho\,{=}\,1.08\,\sigma^{-3}$ (b). The upper wall velocities $U$
are tabulated in the inset. The left vertical axis indicates the
location of the fcc lattice plane at $z\,{=}\,-15.24\,\sigma$. The
dashed line at $z\,{=}\,-14.74\,\sigma$ denotes the reference plane
for computing the slip length.} \label{mol_dens}
\end{figure}

The representative density profiles are shown in
Fig.\,\ref{mol_dens} for $\rho\,{=}\,1.04\,\sigma^{-3}$ and
$\rho\,{=}\,1.08\,\sigma^{-3}$ and two values of the upper wall
speed. The density oscillations extend to a distance of about
$5-6\,\sigma$ from the walls and the profiles are uniform in the
central bulk region. The magnitude of the first peak near the wall
defines a contact density $\rho_c$. The amplitude of the density
oscillations is reduced with increasing upper wall speed. Notice
that at low $U$ in the case $\rho\,{=}\,1.08\,\sigma^{-3}$ shown in
Fig.\,\ref{mol_dens}\,(b), the local density minimum between the
first and second peaks is almost zero. It implies that the fluid
monomers rarely jump between the first two layers, and it is,
therefore, expected that the averaged velocity profiles will have
relatively poor statistics in that region (see below).

Figure\,\ref{velo_P6} shows the averaged velocity profiles in
steady-state flow for the lowest fluid density
($\rho\,{=}\,1.04\,\sigma^{-3}$) considered in this study. The
velocity profiles remain linear throughout the channel except for a
larger slope inside the first fluid layer. A finite slip velocity is
noticeable even at very low upper wall speed
$U\,{=}\,\,0.01\,\sigma/\tau$. The data are noisy because the
averaged velocity component in the $\hat{x}$ direction is much
smaller than the thermal fluid velocity $v^2_{T}\!=k_BT\!/m$. The
velocity profile for $U\,{=}\,\,0.01\,\sigma/\tau$ bends slightly
near the walls, which implies that the interfacial viscosity is
higher than the fluid bulk viscosity. A small curvature in the bulk
region of the velocity profile at $U\,{=}\,\,4\,\sigma/\tau$ might
be related to the nonuniform heating up of the fluid at high shear
rates~\cite{Priezjev08}. The normalized slip velocity increases with
increasing upper wall speed.

The averaged velocity profiles at the higher melt density
($\rho\,{=}\,1.08\,\sigma^{-3}$) are reported in
Fig.\,\ref{velo_P8}. At small values of the upper wall speed, the
slip velocity of the first fluid layer is barely noticeable and the
profiles are highly curved near the walls due to the presence of the
viscous interfacial layer with thickness of about $2-3\,\sigma$. The
statistical fluctuations are relatively large near the walls because
of the pronounced density layering [e.g., see
Fig.\,\ref{mol_dens}\,(b)]. The effective no-slip boundary plane is
located inside the fluid domain at a distance of about $2\,\sigma$
from the wall. With increasing upper wall speed, the fluid velocity
profiles become linear, the no-slip boundary plane is displaced out
of the fluid region, and the slip velocity increases.

% following a thorough equilibration

\begin{figure}[t]
\includegraphics[width=12.cm,angle=0]{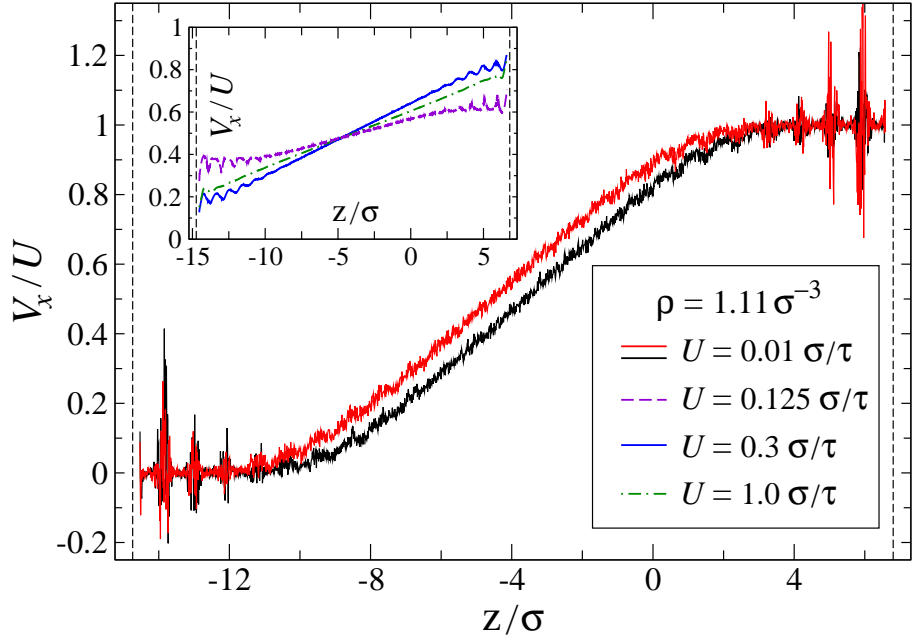}
\caption{(Color online) Average normalized velocity profiles for the
upper wall speed $U\,{=}\,\,0.01\,\sigma/\tau$ (see text for
description). The vertical dashed lines indicate liquid/solid
interfaces at $z\,{=}\,-14.74\,\sigma$ and $z\,{=}\,6.81\,\sigma$.
The inset shows averaged velocity profiles for the upper wall speeds
$U\,{=}\,\,0.125\,\sigma/\tau$ (dashed violet curve),
$U\,{=}\,\,0.3\,\sigma/\tau$ (continuous blue curve), and
$U\,{=}\,\,1.0\,\sigma/\tau$ (dash-dotted green line).}
\label{velo_PX}
\end{figure}

The normalized velocity profiles for the highest fluid density
$\rho\,{=}\,1.11\,\sigma^{-3}$ are plotted in Fig.\,\ref{velo_PX}.
At the lowest upper wall speed $U\,{=}\,\,0.01\,\sigma/\tau$, the
first four monolayers stick to the walls but the velocity profiles
remain linear in the middle of the channel. Due to the low
probability of finding chain segments in between the monolayers the
statistical uncertainties are much larger near the walls than in the
bulk region. We found that the shape of the flow profiles depends on
how the system was prepared. In one case, the upper wall speed was
increased from zero to $U\,{=}\,\,0.01\,\sigma/\tau$ following the
equilibration procedure described in the previous section. The
averaged velocity profile is marked by the lower black curve in
Fig.\,\ref{velo_PX}. In the other case, the upper wall speed was
first increased to $U\,{=}\,\,0.5\,\sigma/\tau$, and then, after the
equilibration period of about $10^5\,\tau$, gradually reduced to
$U\,{=}\,\,0.01\,\sigma/\tau$. The corresponding velocity profile is
shown by the upper red curve in Fig.\,\ref{velo_PX}. Note that the
width of the flowing regions is nearly the same in both cases but
the thickness of the immobile interfacial layers varies from about
four to five molecular diameters. We also found that at higher upper
wall speeds ($0.1\lesssim U\,\tau/\sigma\lesssim 0.25$) the first
fluid layer slides with a finite velocity past the substrate and the
weak oscillations in the velocity profiles near the walls correlate
with the fluid density layering (see inset in Fig.\,\ref{velo_PX}).
Finally, as shown in the inset of Fig.\,\ref{velo_PX}, the velocity
profiles become linear throughout the channel at higher upper wall
speeds ($U\gtrsim0.3\,\sigma/\tau$) and the slip velocity increases
monotonically with increasing $U$.

The slope of the linear part of the velocity profiles in the bulk
region of the channel ($6\,\sigma$ away from the walls) was used to
compute the shear rate, shear stress and slip length. We define the
slip length as a location of the point where linearly extrapolated
velocity profile vanishes. Negative value of the slip length implies
that the effective no-slip boundary plane is displaced into the bulk
fluid domain. The velocity of the first fluid layer with respect to
the lower stationary wall was computed as follows
\begin{equation}
V_{1}=\int_{z_0}^{z_1}\!V_x(z)\rho(z)dz \,\Big/
\int_{z_0}^{z_1}\!\rho(z)dz, \label{velo_defin}
\end{equation}
where the limits of integration ($z_0=-14.50\,\sigma$ and
$z_1=-13.85\,\sigma$) were determined from the width of the first
peak in the density profile. Note that even if the flow profiles are
linear throughout the channel, the velocity of the first layer $V_1$
is slightly larger than the slip velocity computed from the Navier
relation $V_s\,{=}\,\,\dot{\gamma}L_s$ (by about $0.5\,\dot{\gamma}$
in units of $\sigma/\tau$). This is because the slip length is
defined with respect to the reference plane $0.5\,\sigma$ away from
the inner fcc lattice plane and the first fluid layer is located
approximately $\sigma$ away from the fcc plane (see
Fig.\,\ref{mol_dens}).

\subsection{Shear rate dependence of the melt viscosity and slip length}

\begin{figure}[t]
\includegraphics[width=12.cm,angle=0]{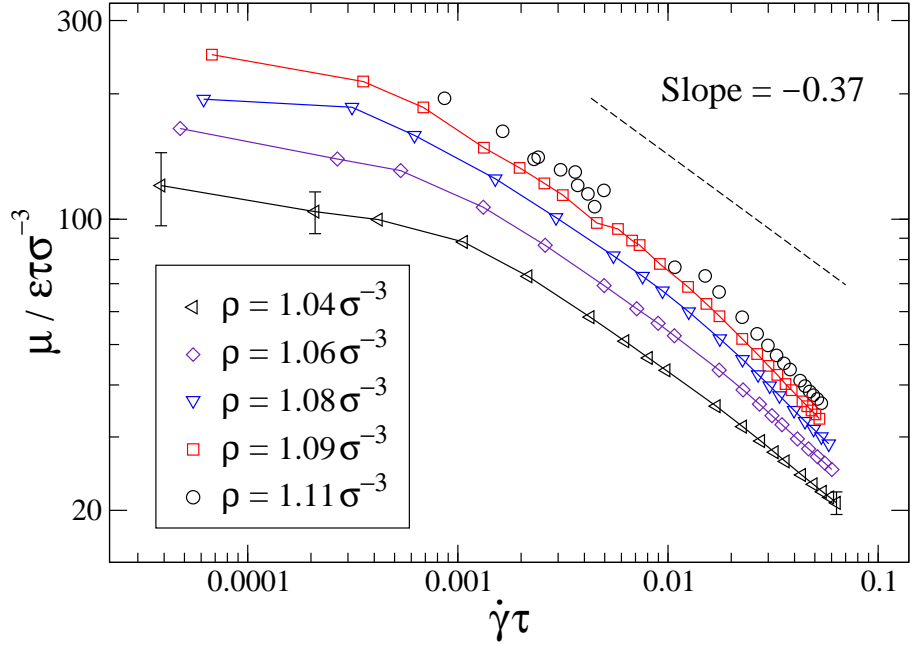}
\caption{(Color online) Shear rate dependence of the bulk viscosity
$\mu/\,\varepsilon\tau\sigma^{-3}$ for the indicated values of the
polymer density. The dashed line with a slope $-0.37$ is shown for
reference. Solid curves are a guide for the eye.}
\label{visc_shear_all}
\end{figure}

We first estimate the polymer melt viscosity which is defined as a
ratio of shear stress to shear rate, i.e.,
$\sigma_{xz}\,{=}\,\,\mu\dot{\gamma}$. The shear stress was computed
using the Kirkwood relation~\cite{Kirkwood} in the bulk region
($6\,\sigma$ away from the confining walls), where the fluid
structure is uniform and the velocity profiles are linear even at
low shear rates. The viscosity is plotted in
Fig.\,\ref{visc_shear_all} as a function of shear rate for the
indicated values of the polymer density. The Newtonian regime is
observed only in a narrow range of shear rates, and it is followed
by the crossover to the shear-thinning behavior, which occurs at
higher shear rates when the melt density is reduced. The dashed line
in Fig.\,\ref{visc_shear_all} corresponds to the power law decay
with the exponent $-0.37$ reported for the lower density polymer
melts ($0.86\leqslant\rho\,\sigma^3\leqslant1.02$) at high shear
rates~\cite{Priezjev08}. The data presented in
Fig.\,\ref{visc_shear_all} indicate that the melt viscosity in the
shear-thinning regime decreases faster at higher fluid densities.
The statistical errors due to thermal fluctuations are relatively
large at low shear rates.

% The data are scattered at $\rho\,{=}\,1.11\,\sigma^{-3}$ due to.

\begin{figure}[t]
\includegraphics[width=12.cm,angle=0]{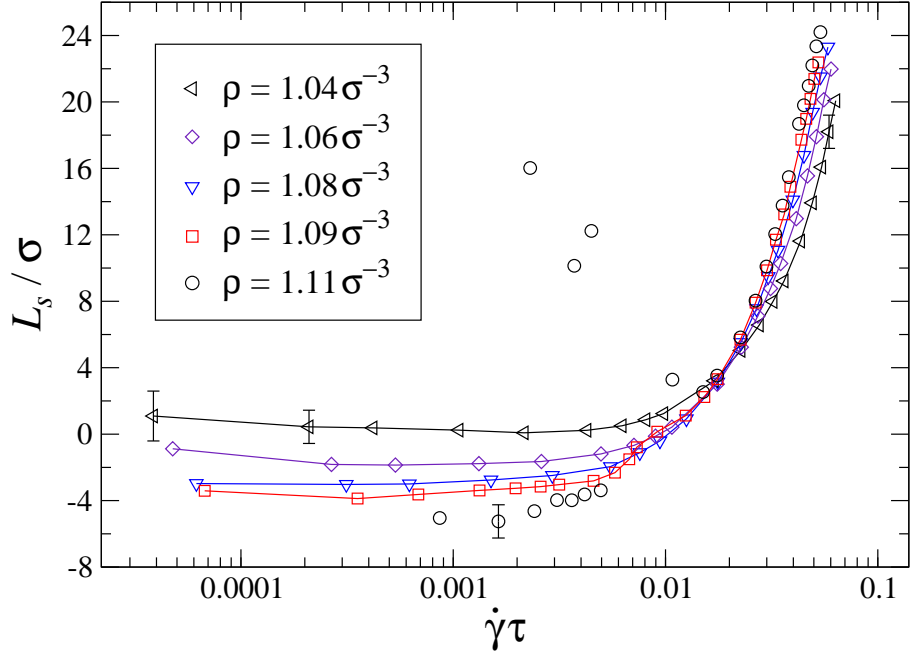}
\caption{(Color online) Slip length $L_s/\sigma$ as a function of
shear rate for the indicated values of the polymer melt density. At
low shear rates and $\rho\,{=}\,1.11\,\sigma^{-3}$ the slip length
is multivalued. Solid curves are drawn to guide the
eye.}\label{shear_ls_all}
\end{figure}

The variation of the slip length as a function of shear rate is
presented in Fig.\,\ref{shear_ls_all} for all melt densities
considered. As expected from the shape of the velocity profiles
described in the previous section, the slip length at low shear
rates is negative (except for $\rho\,{=}\,1.04\,\sigma^{-3}$) and
its magnitude is approximately equal to the thickness of the viscous
interfacial layer. With increasing shear rate, the velocity profiles
become linear, implying that the local viscosity of the boundary
layer is reduced, and the slip length increases rapidly. At the
highest melt density $\rho\,{=}\,1.11\,\sigma^{-3}$ and low shear
rates, the thickness of the interfacial layer depends on the
equilibration procedure and the slip length cannot be uniquely
defined. The uncertainty in the slip length related to the thickness
of the immobile interfacial layer and the slope of the velocity
profile in the bulk region is about $2\,\sigma$. At higher upper
wall speeds ($0.1\lesssim U\,\tau/\sigma\lesssim 0.25$) and
$\rho\,{=}\,1.11\,\sigma^{-3}$, the first fluid layer is sliding
with a finite velocity and the slip length is relatively large.
Multivalued slip lengths were also reported in shear flow of simple
fluids past smooth surfaces with high wall-fluid interaction
energy~\cite{Thompson90}. Finally, we comment that the data
presented in Fig.\,\ref{shear_ls_all} cannot be well fitted by the
power law function proposed in Ref.\,\cite{Nature97}.

% reproduceable bump at $\rho\,{=}\,1.09\,\sigma^{-3}$.

In our simulations, the upper wall speed was varied so that the slip
velocity remains less than about the fluid thermal velocity. We
performed test runs at higher upper wall speeds
($U\gtrsim4\,\sigma/\tau$) and observed a different regime, where
the the slip length becomes a nonmonotonic function of shear rate.
It was also recently shown that the slip length at the interface
between short chain polymers and smooth thermal walls approaches a
constant value at high shear rates~\cite{LichJFM08,LichPRL08}. In
the present study, the behavior of the slip length at very large
slip velocities and high shear rates was not examined in detail.

\begin{figure}[t]
\includegraphics[width=12.cm,angle=0]{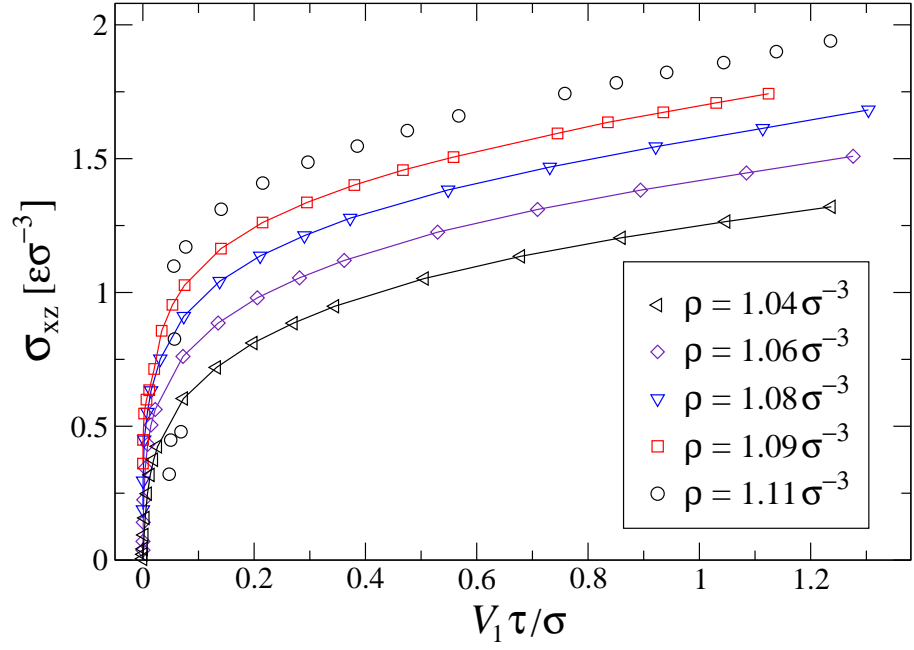}
\caption{(Color online) Shear stress $\sigma_{xz}$ (in units of
$\varepsilon\sigma^{-3}$) averaged in the bulk region as a function
of the slip velocity of the first fluid layer for the indicated
values of the polymer melt density. Solid curves are a guide for the
eye.}\label{stress_velo}
\end{figure}

The rate-dependent boundary conditions can be reformulated in terms
of the wall shear stress and slip velocity. In the steady-state
shear flow, the stress across any plane parallel to the confining
walls is the same, and, therefore, the shear stress computed in the
bulk region is equal to the wall shear stress. The slip velocity of
the first fluid layer was calculated by averaging the velocity
profile over the width of the first density peak using
Eq.\,(\ref{velo_defin}). Note that the slip velocity computed from
the Navier relation ($V_s=\dot{\gamma}L_s$) is smaller than $V_1$ or
can be even negative if the velocity profiles are curved near the
interface [e.g., see Fig.\,\ref{schematic}\,(c) and
Fig.\,\ref{schematic}\,(d)]. The shear stress averaged in the bulk
region is plotted in Fig.\,\ref{stress_velo} as a function of the
slip velocity of the first fluid layer for the indicated polymer
melt densities. The shear stress increases rapidly at small slip
velocities and grows steadily at large $V_1$. At the highest melt
density $\rho\,{=}\,1.11\,\sigma^{-3}$, the time-averaged shear
stress is discontinuous at small slip velocities.

\begin{figure}[t]
\includegraphics[width=12.cm,angle=0]{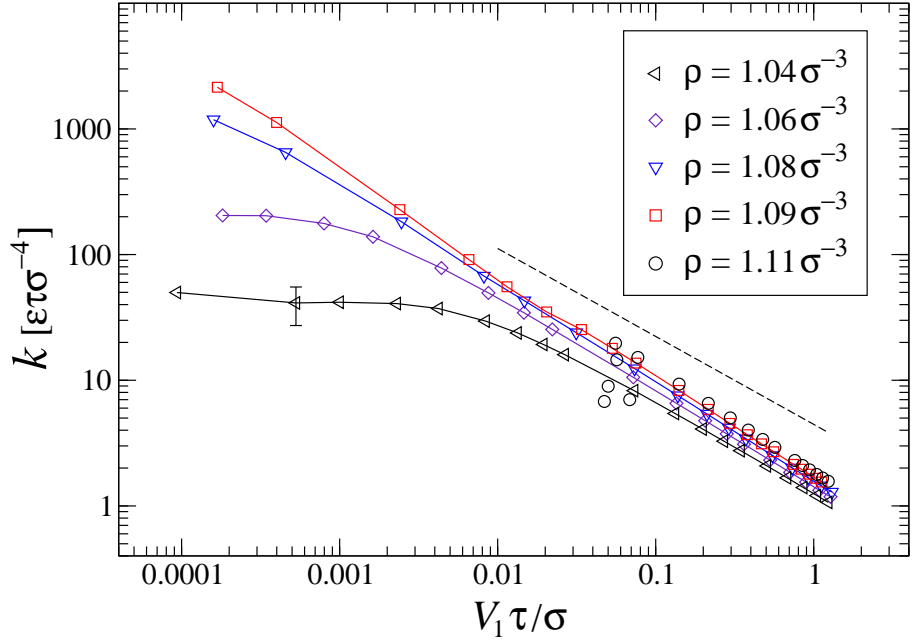}
\caption{(Color online) Log-log plot of the friction coefficient
$k=\sigma_{xz}/V_1$ (in units of $\varepsilon\tau\sigma^{-4}$) as a
function of the slip velocity [computed from
Eq.\,(\ref{velo_defin})] for the tabulated values of the polymer
density. The dashed line with a slope $-0.7$ is plotted for
reference. The same data as in Fig.\,\ref{stress_velo}. Solid curves
are drawn as a guide for the eye.}\label{friction_velo}
\end{figure}

The nonlinear relation between the shear stress and slip velocity
shown in Fig.\,\ref{stress_velo} can be used to determine the
friction coefficient per unit area $\sigma_{xz}\,{=}\,\,kV_1$ as a
function of $V_1$. We comment that the ratio of the shear viscosity
to the slip length cannot be used to compute the friction
coefficient at the liquid/solid interface when the velocity profiles
are curved near the surface and the slip length is extracted from
the bulk part of the profiles. In the previous
study~\cite{Priezjev08}, the simulations were performed at lower
polymer melt densities and the velocity profiles remained linear at
all shear rates examined. In the range of densities
($0.86\leqslant\rho\,\sigma^3\leqslant1.02$), the friction
coefficient ($k=\mu/L_s$) as a function of the slip velocity was
well described by the following equation
\begin{equation}
k/k^{\ast}=[1+(V_s/V_s^{\ast})^2]^{-0.35}, \label{friction_law}
\end{equation}
where $k^{\ast}$ and $V_s^{\ast}$ are the normalization
parameters~\cite{Priezjev08}. In the present study, the friction
coefficient is plotted as a function of the slip velocity in
Fig.\,\ref{friction_velo}. The data can be well fitted by the
empirical formula Eq.\,(\ref{friction_law}) at lower melt densities
$\rho\,{\leqslant}\,1.06\,\sigma^{-3}$ (not shown). At higher melt
densities $\rho\,{\gtrsim}\,1.08\,\sigma^{-3}$, the plateau regime,
where the friction coefficient is independent of the slip velocity,
is absent and the slope of the power law decay is slightly smaller
than $-0.7$ (see the dashed line in Fig.\,\ref{friction_velo}).

%%% comment on stick-slip    minimal slip velocities we could definitely resolve

\subsection{Friction coefficient and fluid structure in the first layer}

The connection between friction at the liquid/solid interface and
fluid structure induced by the periodic surface potential was
established for monatomic
fluids~\cite{Thompson90,Barrat99fd,Priezjev05,Priezjev06,Priezjev07,PriezjevJCP},
polymer melts~\cite{Priezjev04,Priezjev08}, adsorbed
monolayers~\cite{Smith96,Tomassone97} and adsorbed polymer
layers~\cite{Muser06}. The in-plane structure factor in the first
fluid layer near the solid substrate is defined as
\begin{equation}
S(\mathbf{k})=\frac{1}{N_{\ell}}\,\,\Big|\sum_{j=1}^{N_{\ell}}
e^{i\,\mathbf{k}\cdot\mathbf{r}_j}\Big|^2,
\end{equation}
where $\mathbf{r}_j\,{=}\,(x_j,y_j)$ is the position vector of the
$j$-th monomer and $N_{\ell}$ is the number of monomers within the
layer~\cite{Thompson90}. Typically, the structure factor exhibits a
circular ridge at the wavevector $|\mathbf{k}|\approx2\pi\!/\sigma$
due to short range ordering of the fluid monomers. In addition, at
sufficiently high wall-fluid interaction energy, several sharp peaks
appear in the structure factor at the reciprocal lattice vectors of
the crystal wall. It is well established that the magnitude of the
peak at the first reciprocal lattice vector in the shear flow
direction correlates well with the friction coefficient at the
interface between a simple LJ liquid and a solid wall composed out
of periodically arranged LJ
atoms~\cite{Thompson90,Barrat99fd,Priezjev07}.

\begin{figure}[t]
\includegraphics[width=12.cm,angle=0]{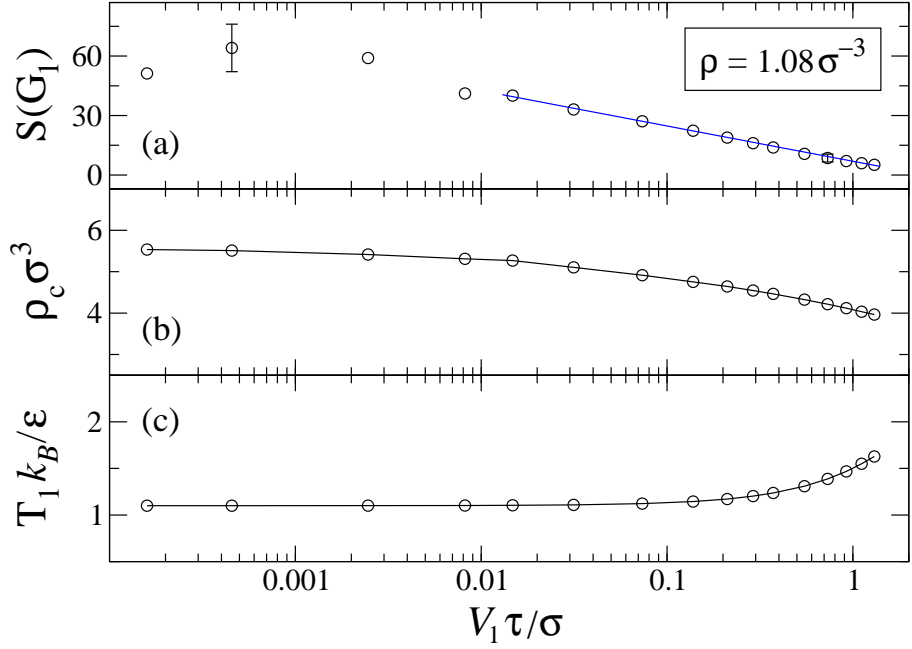}
\caption{(Color online) Structure factor $S(\mathbf{G}_1)$ evaluated
at the first reciprocal lattice vector
$\mathbf{G}_1\,{=}\,(7.23\,\sigma^{-1},0)$ in the shear flow
direction (a), contact density near the stationary lower wall (b),
and temperature (c) of the first fluid layer as a function of the
slip velocity $V_1$ (in units of $\sigma/\tau$). The solid blue line
$y=6.93-7.73\,\text{ln}(x)$ represents the best fit to the data.
Solid curves are a guide for the eye.}\label{velo_P8_T_rho_sk}
\end{figure}

In the present study, the first reciprocal lattice vector in the
shear flow direction $\mathbf{G}_1\,{=}\,(7.23\,\sigma^{-1},0)$ is
slightly displaced from the wavevector
$|\mathbf{k}|\approx2\pi\!/\sigma$. The averaged structure factor
$S(\mathbf{G}_1)$ is plotted in Fig.\,\ref{velo_P8_T_rho_sk}\,(a) as
a function of the slip velocity of the first fluid layer for the
polymer density $\rho\,{=}\,1.08\,\sigma^{-3}$. The error bars are
relatively large at small slip velocities due to the slow relaxation
dynamics of the polymer chains in the interfacial layer (see next
section). At higher slip velocities, fluid monomers spend less time
in the minima of the periodic surface potential, and, as shown in
Fig.\,\ref{velo_P8_T_rho_sk}\,(a), the magnitude of the
surface-induced peak in the structure factor decreases
logarithmically with increasing $V_1$. Furthermore, as expected from
the density profiles shown in Fig.\,\ref{mol_dens} for different
upper wall speeds, the contact density is reduced at higher slip
velocities [see Fig.\,\ref{velo_P8_T_rho_sk}\,(b)]. Similarly to the
definition of the slip velocity given by Eq.\,(\ref{velo_defin}),
the temperature of the first fluid layer was computed as follows
\begin{equation}
T_{1}=\int_{z_0}^{z_1}\!T(z)\rho(z)dz \,\Big/
\int_{z_0}^{z_1}\!\rho(z)dz, \label{temper_defin}
\end{equation}
where $T(z)$ is the local kinetic temperature and the limits of
integration ($z_0=-14.50\,\sigma$ and $z_1=-13.85\,\sigma$) were
determined from the width of the first density peak. The variation
of the monolayer temperature as a function of the slip velocity is
presented in Fig.\,\ref{velo_P8_T_rho_sk}\,(c). At small slip
velocities $V_1\lesssim0.03\,\sigma/\tau$, the temperature is equal
to value $T\,{=}\,1.1\,\varepsilon/k_B$ set by the Langevin
thermostat. With increasing upper wall speed, the temperature of the
first fluid layer gradually rises up to
$T\,{\approx}\,1.6\,\varepsilon/k_B$ at the highest slip velocity
$V_1\approx1.3\,\sigma/\tau$ reported in
Fig.\,\ref{velo_P8_T_rho_sk}\,(c). At shear rates
$\dot{\gamma}\gtrsim0.01\,\tau^{-1}$, the temperature profiles
across the channel $T(z)$ become nonuniform and the heating up is
larger near the interfaces~\cite{Priezjev08}.

\begin{figure}[t]
\includegraphics[width=12.cm,angle=0]{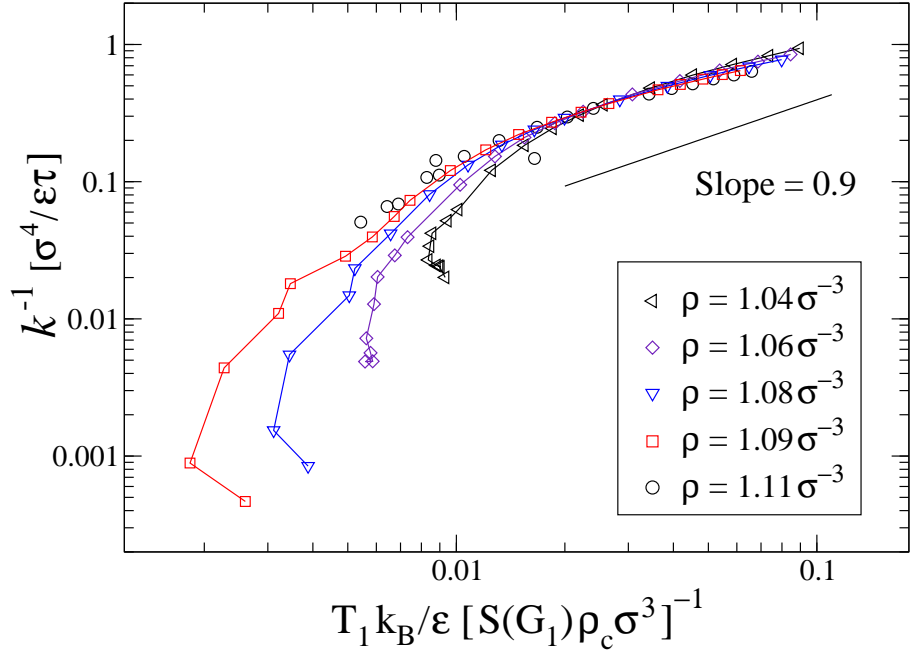}
\caption{(Color online) Log-log plot of the inverse friction
coefficient $k^{-1}=V_1/\sigma_{xz}$ (in units of
$\sigma^4/\varepsilon\tau$) as a function of
$T_1k_B/\varepsilon\,[S(\mathbf{G}_1)\,\rho_c\,\sigma^3]^{-1}$
evaluated in the first fluid layer for the indicated polymer melt
densities. The solid line with a slope $0.9$ is plotted as a
reference.} \label{inv_fr_vs_T_div_S7_ro_c_low}
\end{figure}

\begin{figure}[t]
\includegraphics[width=12.cm,angle=0]{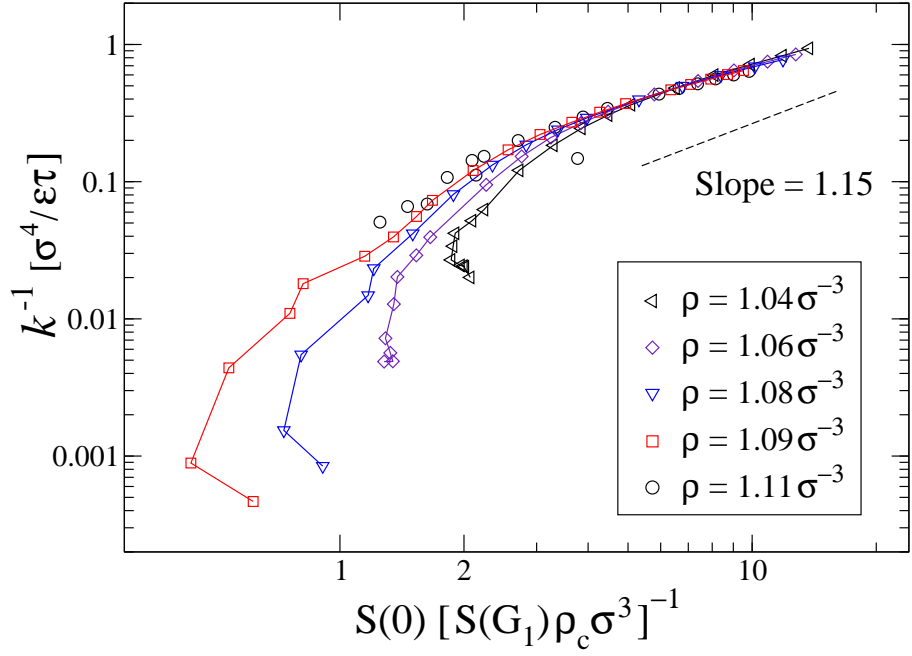}
\caption{(Color online) Log-log plot of the inverse friction
coefficient $k^{-1}=V_1/\sigma_{xz}$ (in units of
$\sigma^4/\varepsilon\tau$) as a function of the variable
$S(0)/\,[S(\mathbf{G}_1)\,\rho_c\,\sigma^3]$ computed in the first
fluid layer. The fluid densities are listed in the inset. The dashed
line with a slope $1.15$ is shown for reference.}
\label{inv_fr_vs_S0_div_S7_ro_c_low}
\end{figure}

The dependence of the friction coefficient on the structure factor,
contact density and temperature of the first fluid layer was studied
in the recent paper~\cite{Priezjev08} at lower polymer melt
densities ($0.86\leqslant\rho\,\sigma^3\leqslant1.02$). Except for
the densities $\rho\,{=}\,1.00\,\sigma^{-3}$ and $1.02\,\sigma^{-3}$
at low shear rates, the data for the friction coefficient
($k=\mu/L_s$) were found to collapse on master curves when plotted
as a function of either $T_1/[S(\mathbf{G}_1)\,\rho_c]$ or
$S(0)/\,[S(\mathbf{G}_1)\,\rho_c]$. In both cases, the data could be
well fitted by a power law function with the exponents $-0.9$ and
$-1.15$, respectively~\cite{Priezjev08}. At higher polymer densities
($1.04\leqslant\rho\,\sigma^3\leqslant1.11$) considered in the
present study, the inverse friction coefficient is plotted in
Figures \ref{inv_fr_vs_T_div_S7_ro_c_low} and
\ref{inv_fr_vs_S0_div_S7_ro_c_low} as a function of the combined
variables $T_1/[S(\mathbf{G}_1)\,\rho_c]$ and
$S(0)/\,[S(\mathbf{G}_1)\,\rho_c]$, respectively. The collapse of
the data holds at small values of the friction coefficient
$k\lesssim4\,\varepsilon\tau\sigma^{-4}$ and the surface-induced
peak in the structure factor $S(\mathbf{G}_1)\lesssim14$. Thus, the
simulations at higher melt densities provide an upper bound for the
friction coefficient ($k\approx4\,\varepsilon\tau\sigma^{-4}$),
below which the data are described by a single master
curve~\cite{Priezjev08}. These results support the conclusion from
the previous studies that at the interface between an atomically
smooth solid wall and a simple~\cite{Priezjev07} or
polymeric~\cite{Priezjev08} fluid, the friction coefficient is
determined by a combination of parameters evaluated in the first
fluid layer.

\subsection{Relaxation dynamics of polymer chains}

Equilibrium molecular dynamics studies of polymer melts confined
between attractive walls have shown that the relaxation of polymer
chains slows down near the interfaces and becomes bulk-like at
distances of about two radiuses of gyration away from the
walls~\cite{Bitsanis93,Doi01}. In order to probe the relaxation
dynamics of polymer chains in shear flow we evaluated the
autocorrelation function of the normal modes in the direction
perpendicular to the plane of shear. The $\hat{y}$ component of the
normal coordinates for a discrete polymer chain of $N$ monomers is
given by
\begin{equation}
\text{Y}_p(t)=\frac{1}{N}\sum_{i=1}^N
y_i(t)\,\textrm{cos}\,\frac{p\pi(i-1)}{N-1}, \label{normal_mode_y}
\end{equation}
where $y_i$ is the component of the position vector of the $i$-th
monomer in the chain and $p=0, 1,..., N-1$ is the mode
number~\cite{Binder95}. The longest relaxation time of a polymer
chain is associated with the first mode $p=1$. The normalized time
autocorrelation function for the first normal mode is computed as
follows
\begin{equation}
C_1(t)=\langle \text{Y}_1(t)\cdot \text{Y}_1(0)\rangle/ \langle
\text{Y}_1(0)\cdot \text{Y}_1(0)\rangle. \label{auto_corr_y}
\end{equation}
The relaxation dynamics in heterogeneous systems is usually
described by the stretched exponential (or
Kohlrausch-Williams-Watts) function
$C_1(t)=\textrm{exp}[-(t/\tau_1)^{\beta}]$. The time integral of the
stretched exponential defines the characteristic decay time
$\tau^{\ast}_y=\tau_1\Gamma(\beta^{-1})/\beta$, where $\Gamma$ is
the gamma function and $\beta$ is the stretched exponential
coefficient.

The autocorrelation function, Eq.\,(\ref{auto_corr_y}), was computed
separately in the bulk region ($6\,\sigma$ away from the solid
walls) and inside the interfacial layers (within $3\,\sigma$ from
the walls). During the simulation, the normal coordinate and the
position of the center of mass of each polymer chain were calculated
every $100\,\tau$, and the autocorrelation function was updated only
if the position of the center of mass was inside either bulk or
interfacial regions. The correlation function was averaged for at
least $6\times10^5\tau$ at low shear rates to resolve very slow
dynamics of the polymer chains near the interfaces.

\begin{figure}[t]
\includegraphics[width=12.cm,angle=0]{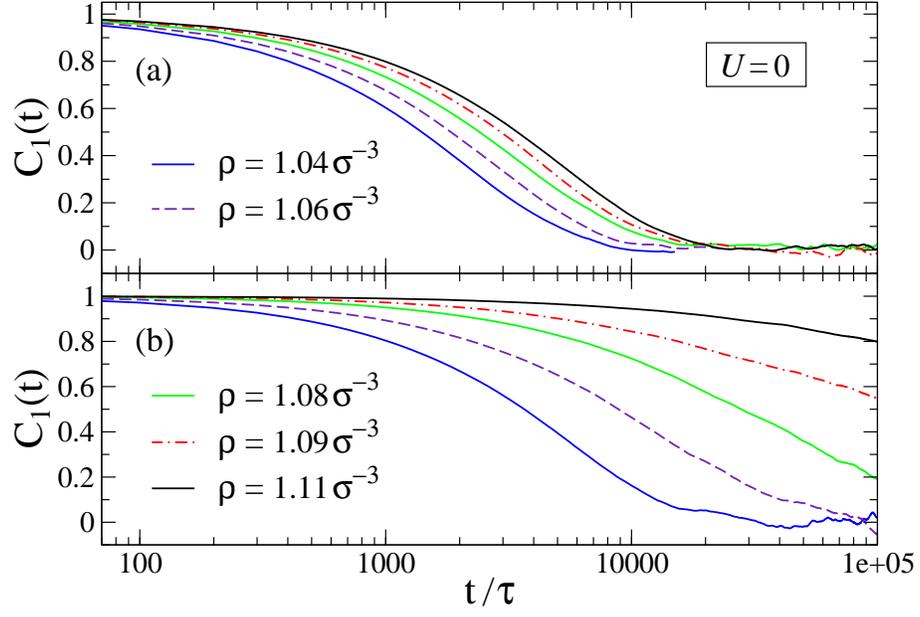}
\caption{(Color online) Normalized time autocorrelation function of
the first normal mode at equilibrium for polymer chains in the bulk
region (a) and near the walls (b) for the indicated values of the
fluid density.} \label{auto_bulk_walls}
\end{figure}

Figure\,\ref{auto_bulk_walls} shows the time autocorrelation
function computed at equilibrium conditions (i.e., $U=0$) for the
indicated polymer densities. As expected, the relaxation of the
polymer chains in the bulk and near the walls is slower at higher
melt densities. The inverse bulk relaxation time, estimated roughly
from $C_1(t)=1/e$ in Fig.\,\ref{auto_bulk_walls}\,(a), correlates
well with the onset of shear thinning of the fluid viscosity
reported in Fig.\,\ref{visc_shear_all}. For each density, the decay
of the correlation function is slower for the chains in the
interfacial layer than in the bulk of the channel. The difference in
relaxation times is especially evident at higher melt densities,
$\rho\geqslant1.09\,\sigma^{-3}$, where only the early stage of
relaxation is reported in Fig.\,\ref{auto_bulk_walls}\,(b). At the
lowest polymer density $\rho\,{=}\,1.04\,\sigma^{-3}$, the decay
time of the interfacial chains is about $2-3$ times larger than in
the bulk, which agrees with the conclusion drawn from the shape of
the velocity profile for $U\,{=}\,\,0.01\,\sigma/\tau$ in
Fig.\,\ref{velo_P6} that the interfacial viscosity is higher than
the bulk value.

\begin{figure}[t]
\includegraphics[width=12.cm,angle=0]{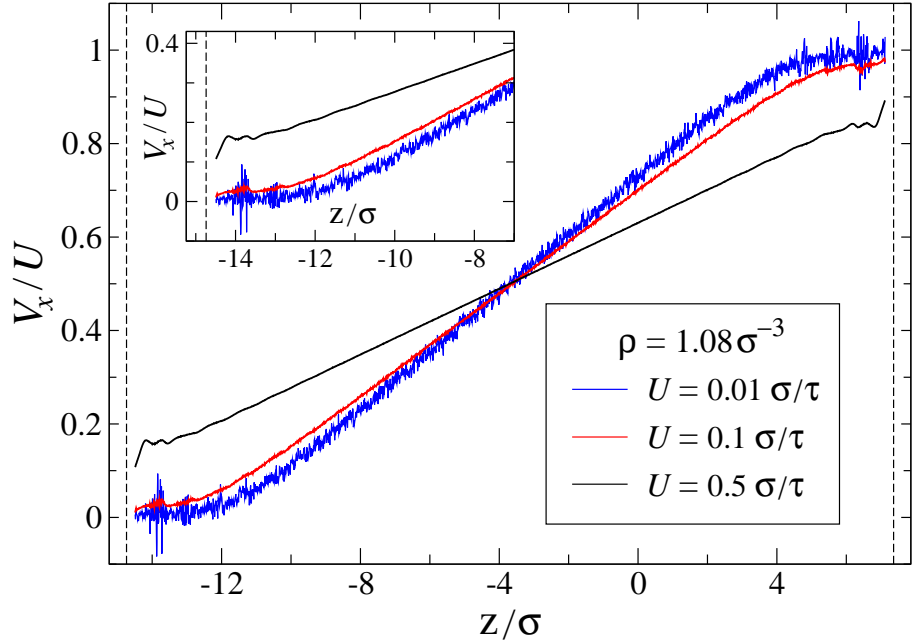}
\caption{(Color online) Averaged velocity profiles for the indicated
upper wall speeds $U$ (in units of $\sigma/\tau$) and the fluid
density $\rho\,{=}\,1.08\,\sigma^{-3}$. The vertical axes indicate
the location of the fcc lattice planes at $z\,{=}\,-15.24\,\sigma$
and $z\,{=}\,7.86\,\sigma$. The dashed lines denote the position of
the liquid/solid interfaces $0.5\,\sigma$ away from the fcc planes.
The region near the lower wall is enlarged in the inset.}
\label{velo_P8}
\end{figure}

\begin{figure}[t]
\includegraphics[width=12.cm,angle=0]{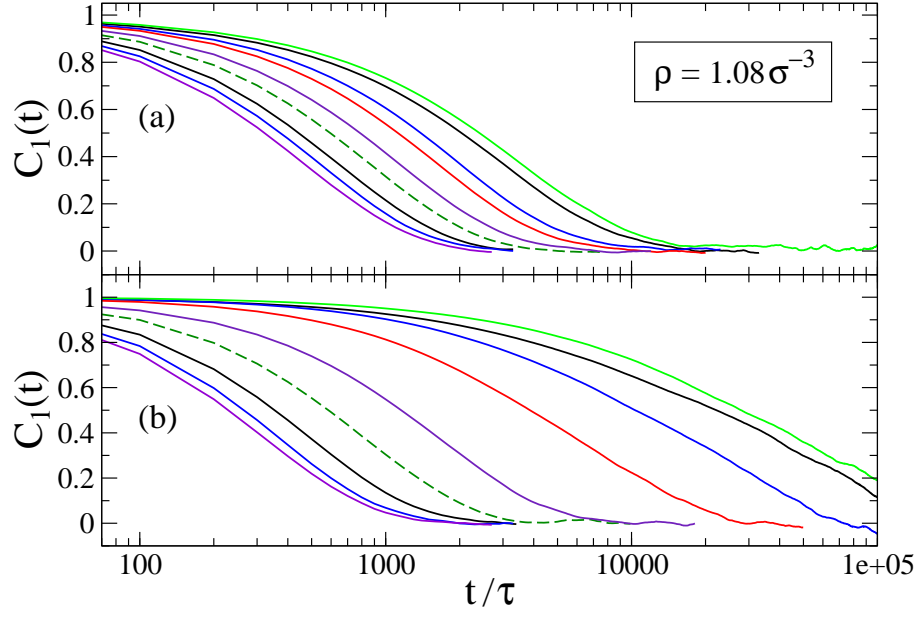}
\caption{(Color online) Time autocorrelation function of the first
normal mode for polymer chains in the bulk region (a) and near the
walls (b) and the fluid density $\rho\,{=}\,1.08\,\sigma^{-3}$. The
upper wall speed from right to left is
$U=0,\,0.025,\,0.1,\,0.2,\,0.5,\,1.0,\,2.0,\,3.0~\text{and}~4.0$ (in
units of $\sigma/\tau$).} \label{auto_P8_walls}
\end{figure}

The influence of the shear flow on the time autocorrelation function
is demonstrated in Fig.\,\ref{auto_P8_walls} for the polymer density
$\rho\,{=}\,1.08\,\sigma^{-3}$. At low upper wall speeds
$U\lesssim0.1\,\sigma/\tau$, the relaxation dynamics near the
interface is much slower than in the bulk region, implying the
formation of a highly viscous interfacial layer. The spatial
variation of the shear viscosity correlates with the nonlinearity in
the velocity profiles shown in Fig.\,\ref{velo_P8}. When
$U\,{=}\,\,1.0\,\sigma/\tau$, the decay time of the correlation
function is nearly the same across the channel (see the dashed
curves in Fig.\,\ref{auto_P8_walls}) and the velocity profile is
linear (shown for $U\,{=}\,\,0.5\,\sigma/\tau$ in
Fig.\,\ref{velo_P8}). At higher upper wall speeds, the polymer
chains relax slightly faster near the walls than in the bulk,
possibly because of the heating up of the fluid near the walls.

\begin{figure}[t]
\includegraphics[width=12.cm,angle=0]{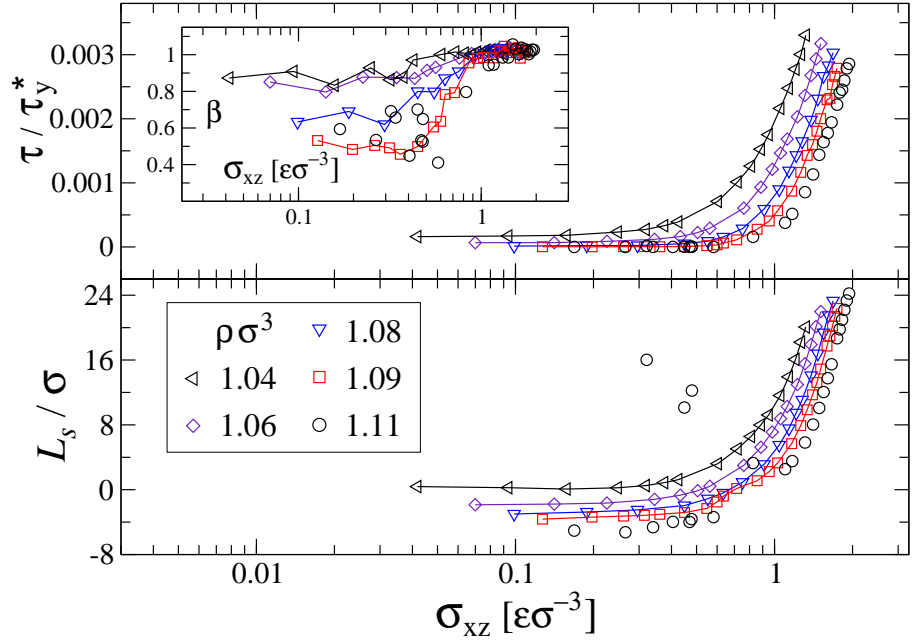}
\caption{(Color online) Inverse relaxation time of the polymer
chains near the walls (top) and the slip length (bottom) as a
function of shear stress $\sigma_{xz}$ (in units of
$\varepsilon\sigma^{-3}$) for the indicated values of the melt
density. The inset shows the stretched exponential coefficient for
the same data.} \label{auto_time_Ls_stress}
\end{figure}

Finally, the relaxation time of the polymer chains near the walls
and the slip length are summarized in
Fig.\,\ref{auto_time_Ls_stress} as a function of shear stress and
polymer density. The leftmost points of the curves shown in
Fig.\,\ref{auto_time_Ls_stress} correspond to the upper wall speed
$U\,{=}\,\,0.01\,\sigma/\tau$. At low shear stress, the relaxation
time varies widely from $\tau^{\ast}_y\approx6.2\times10^2\,\tau$ at
$\rho\,{=}\,1.04\,\sigma^{-3}$ to
$\tau^{\ast}_y\approx3.3\times10^6\,\tau$ at
$\rho\,{=}\,1.11\,\sigma^{-3}$, indicating the presence of a highly
viscous boundary layer at higher melt densities. With increasing
shear stress, the decay time decreases, and, when
$U\gtrsim1.5\,\sigma/\tau$, the relaxation of the polymer chains
near the walls becomes even slightly faster than in the bulk region
(see also Fig.\,\ref{auto_P8_walls}). The dependence the inverse
relaxation time on the shear stress exhibits qualitatively similar
behavior to the slip length, which supports the conclusion (drawn
from the shape of the velocity profiles) that the transition to slip
flow is associated with shear-melting of the boundary layer. The
inset in Fig.\,\ref{auto_time_Ls_stress} shows the variation of the
stretched exponential coefficient as a function shear stress. The
data are scattered at low shear stress due to the slow relaxation of
the polymer chains, and the decay of the autocorrelation function
becomes nearly exponential at higher shear stress.

\section{Conclusions}

In this paper, the rate-dependence of the slip length at the
interface between a dense polymer melt and weakly attractive smooth
walls was studied using molecular dynamics simulations. The melt was
modeled as a collection of linear chain polymers ($N=20$). It was
shown that at low shear rates the velocity profiles are curved near
the wall due to the formation of a highly viscous interfacial layer
and the effective slip length is negative and almost
rate-independent. With increasing upper wall speed, the gradual
transition to steady-state slip flow is associated with
shear-melting of the interfacial layer. The relaxation dynamics of
polymer chains in shear flow was analyzed by evaluating the decay of
time autocorrelation function of the first normal mode in the
vorticity direction. We found that the rate behavior of the slip
length correlates well with the inverse relaxation time of the
polymer chains in the interfacial layer.

The rate-dependent slip boundary conditions were also reformulated
in terms of the friction coefficient at the polymer/wall interface
and slip velocity of the first fluid layer. In agreement with the
results of the previous study~\cite{Priezjev08}, we found that the
friction coefficient at lower melt densities undergoes a transition
from a constant value to the power law decay as a function of the
slip velocity. At higher melt densities the friction coefficient
decays as a power law function in a wide range of slip velocities.
When the magnitude of the surface induced peak in the fluid
structure factor is below a certain value, the friction coefficient
is determined by a combination of parameters (structure factor,
temperature, and contact density) of the first fluid layer near the
solid wall.

\section*{Acknowledgments}

Financial support from the Petroleum Research Fund of the American
Chemical Society is gratefully acknowledged. Computational work in
support of this research was performed at Michigan State
University's High Performance Computing Facility.

\begin{figure}[t]
\includegraphics[width=12.cm,angle=0]{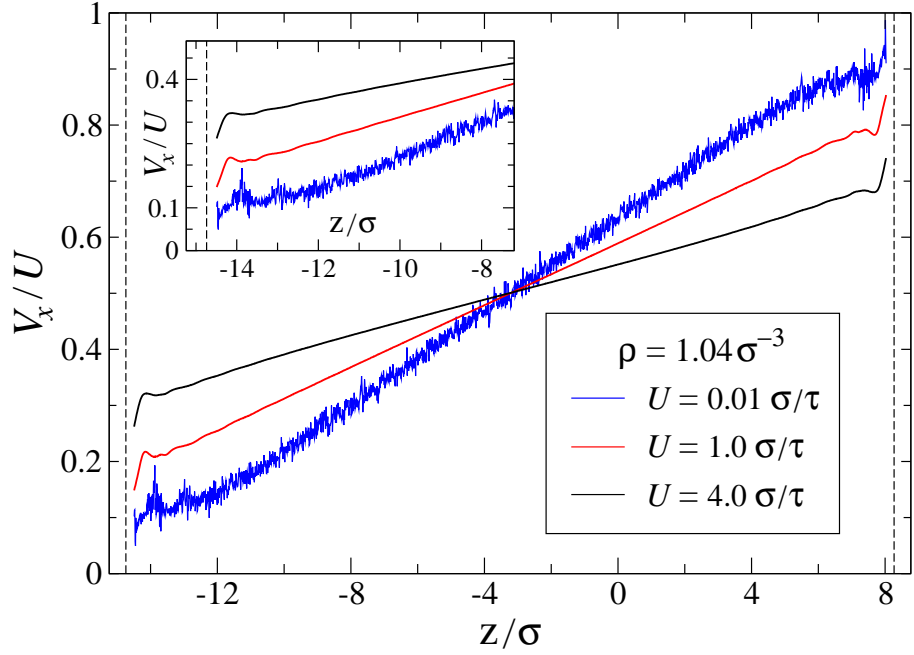}
\caption{(Color online) Average normalized velocity profiles for the
indicated values of the upper wall speed and the fluid density
$\rho\,{=}\,1.04\,\sigma^{-3}$. The vertical axes coincide with the
position of the fcc lattice planes at $z\,{=}\,-15.24\,\sigma$ and
$z\,{=}\,8.77\,\sigma$. The dashed lines denote liquid/solid
interfaces at $z\,{=}\,-14.74\,\sigma$ and $z\,{=}\,8.27\,\sigma$.
The inset shows an enlarged view of the velocity profiles near the
stationary lower wall.} \label{velo_P6}
\end{figure}

\bibliographystyle{prsty}

\end{document}